\documentclass[12pt]{article}
\usepackage[margin=1in, paperwidth=8.5in, paperheight=11in]{geometry}
\usepackage{amsmath}
\usepackage{graphicx}%
\usepackage{amsfonts}%
\usepackage{amssymb}
\usepackage{color}
\usepackage{float,subfigure}
\usepackage[round]{natbib}                  %citations are in ()
\usepackage{cite}
\usepackage{setspace}
\usepackage{comment}
\usepackage{url}
\usepackage{enumerate}
\usepackage{verbatim}
\usepackage{hyperref}
\hypersetup{pdfborder = {0 0 0}} %do not show border
\hypersetup{pdfpagemode=UseNone}

\usepackage{caption} 
\usepackage{float}

\newcommand{\given}{\,\vert\,}

\begin{document}
%\begin{doublespace}
\thispagestyle{empty}
\setcounter{page}{0}

\begin{center}
\singlespacing\textbf{\Large A Two-stage Bayesian Model for Assessing the Geography of Racialized Economic Segregation and Premature Mortality Across US Counties
}\\ 
\singlespacing\textbf{Yang Xu, and Loni Philip Tabb}\\
Department of Epidemiology and Biostatistics, Drexel University,\\ Philadelphia, PA 19104
\end{center}
%\begin{center}
%\singlespacing\textbf{Yang Xu, and Loni Philip Tabb}\\
%Department of Epidemiology and Biostatistics, Drexel University, Philadelphia, PA %19104
%\end{center}

\textsc{Abstract.} Racialized economic segregation, a key metric that simultaneously accounts for spatial, social and income polarization, has been linked to adverse health outcomes, including morbidity and mortality; however, statistical methods for measuring the association between racialized economic segregation and health outcomes are not well-developed and are usually studied at the individual level. In this paper we propose a two-stage Bayesian statistical framework that provides a broad, flexible approach to studying the spatially varying association between premature mortality and racialized economic segregation, while accounting for neighborhood-level latent health factors across US counties. We apply our method by using data from three sources: (1) the CDC WONDER, (2) the County Health Rankings, and (3) the Public Health Disparities Geocoding Project. Findings from our study show that the posterior estimates of latent health factors clearly demonstrate geographical patterning across US counties. Additionally, our results highlight the importance of accounting for the presence of spatial autocorrelation in racialized economic segregation measures, in health equity focused settings.


\newpage
\section{Introduction}
Examining the relationship between residential segregation, a marker of structural racism, and health outcomes has been an active research topic for public health researchers within the past twenty years. A large body of research has shown that residential segregation is not only associated with negative health outcomes including mortality and morbidity, but it also harms the health and well-being of racial and ethnic minorities and society overall \citep{white2011racial}. Residential segregation indices, such as dissimilarity, isolation, and concentration indices, are the most commonly used indicators of structural racism because they reveal the degrees of different environments for the segregated group and the rest of the community \citep{massey1987trends, acevedo2000residential}. These indices typically only capture one aspect of segregation, such as racial segregation. However, given that structural racism is a multidimensional measure and that income disparities have been shown to be an integral aspect of it, there is a great need for a metric that captures both racial and income segregation simultaneously \citep{bailey2017structural}. 

One residential segregation metric that captures both racial and income segregation is the Index of Concentration at the Extremes (ICE). \citet{massey2001residential} developed this metric to measure spatial social polarization of both deprived and privileged socioeconomic groups in one measure, which has been extensively studied in social science. \citet{krieger2016public} further expanded this measure to incorporate race and the simultaneous consideration of race and income, enabling the distinction between extremely low and high concentrations of racial and income differences using a single measure. The ICE measure, specifically ICE$_{race+income}$, which is considered as an operationalization for racialized economic segregation, has recently been applied in population health studies, with a particular focus on social epidemiology. For instance, it has been shown to have an association with hypertension, premature mortality, cancer, COVID19, and cardiovascular health \citep{feldman2015spatial,krieger2016monitoring, krieger2018cancer, chen2021revealing, tabb2022spatially}.
%The Index of Concentration at the Extreme (ICE) is a potential example of such a measure. It was first developed by \citet{massey2001residential} %Massey (2001) 
%for assessing spatial social polarizations of both deprived and privileged socioeconomic groups using a single metric, which has been extensively researched in social sciences. \citet{krieger2016public} %Krieger et al. (2015) 
%further extended ICE to incorporate race, income and race + income interaction (ICE$_{race+income}$) to quantify the extreme low and high concentrations of racial and income differences in a single measure to monitor public health. The ICE measure, specifically ICE$_{ race+income}$, which serves as a proxy for quantifying racialized economic segregation, has recently been applied in population health studies, with a focus on social epidemiology. For instance, it has been shown to have association with hypertension, premature mortality, cancer, COVID19, and cardiovascular health \citep{feldman2015spatial,krieger2016metrics,krieger2018cancer,chen2021revealing,tabb2022spatially}.

%The ICE measure has been utilized more recently in population health studies, with a particular focus on social epidemiology. For instance, it has been shown to have association with hypertension, premature mortality, cancer, COVID19, and cardiovascular health \citep{feldman2015spatial,krieger2016metrics,krieger2018cancer,chen2021revealing,tabb2022spatially}.
%(Feldman et al., 2015; Krieger et al., 2015, 2018; Chen and Krieger, 2021; Tabb et al., 2021). 
\par
Statistical methods for measuring racialized economic segregation and health outcomes relationship are not well-developed and are usually studied at the individual level. The majority of studies in this field typically consider racialized economic segregation as a fixed effect for health outcomes in a given study region, whereas, in fact, health outcomes in each region may spatially vary or cluster because of differences in the exposure level of racialized economic segregation. For instance, it is highly likely that neighboring regions have a similar racialized economic segregation level, indicating spatial correlation in the data. Ignoring such an effect may not help identify the underlying spatial relationship between racialized economic segregation and health outcomes. In addition, the focus of these studies is often on measuring covariates at the individual-level; however, the neighborhood-level characteristics have not been adequately evaluated in this context. Numerous studies have found that neighborhood-level characteristics such as health behavior, social and economic conditions, and the physical environment are essential contributors to health. However, these contributors, are considered latent factors, which means they cannot be directly observed and must be carefully evaluated in order to be constructed. Additionally, it has been challenging to construct these latent components in a way that is both suitable and effective in including spatial effects.

To our knowledge, assessing the spatially varying relationship between racialized economic segregation and health outcomes while controlling for geographical-level latent health factors has not been studied. Thus, we propose a two-stage Bayesian statistical framework to achieve this goal. In stage 1, a Bayesian spatial factor model is utilized to estimate four latent health factors based on the County Health Rankings (CHRs) conceptual model while appropriately accounting for the hypothesized spatial correlation present in the data. In stage 2, we examine the relationship between health outcomes and varying racialized economic segregation by introducing a spatially varying coefficient model and adjusting for the latent health factors. We choose premature mortality as our health outcome.

Methods and applications of Bayesian spatial factor models have varied in the past, including multiple and single latent factors, continuous and discrete data, adding temporal effects, and using models for predictions \citep{wang2003generalized,hogan2004bayesian,abellan2007bayesian,tzala2008bayesian,nethery2015common}. Spatially varying coefficient models have also been widely applied and studied in many fields, including econometrics, ecology, environmental science, and epidemiology \citep{nakaya2005geographically,bitter2007incorporating,waller2007quantifying,finley2011comparing,tabb2022spatially}.
%(Wang and Wall, 2003; Hogan and Tchernis, 2004; \citet{abellan2007bayesian}, 2007; Tzala and Best, 2007; Nethery et al., 2015). Spatially varying coefficient models have also been widely applied and studied in many fields, including econometrics, ecology, environmental science, and epidemiology \citep{nakaya2005geographically,bitter2007incorporating,waller2007quantifying,finley2011comparing,tabb2022spatially}.
%Building on these prior studies, out first stage model applies elements of Bayesian spatial factor model in order to properly assess the latent health factors from the CHRs framework. 
%Spatially varying coefficient models have also been widely applied and studied in many fields, including econometrics, ecology, environmental science, and epidemiology \citep{nakaya2005geographically,bitter2007incorporating,waller2007quantifying,finley2011comparing,tabb2022spatially}.
%(Nakaya et al.,2005; Bitter et al.,2007; Waller et al., 2007; Finley, 2011; Tabb et al., 2022). 
%Based on these studies, we apply the spatially varying coefficient in the second stage to analyze the spatial relationship between racialized economic segregation and premature mortality and extend this model by including estimated neighborhood-level latent health factors.
Building on these prior studies, our proposed modeling framework could adequately reduce the dimension of spatially correlated data, lead to more efficient computation in assessing association without loss of information, and reveal the underlying relationship between health outcome and varying racialized economic segregation.

This manuscript is organized as follows: Section 2 describes the dataset that motivates our method. Section 3 provides the details of the two-stage Bayesian framework. We then provide the results in Section 4 and close with a discussion in Section 5.

\section{Data}
This analysis is conducted for 3108 counties in the contiguous US. We aim to quantify the spatially varying impact of racialized economic segregation on premature mortality, as well as accounting for health factors. Premature mortality (death before 75 years old) data for each county were obtained from the CDC WONDER between 2014 and 2018. The observed number of deaths ($Y$) depend on the population size and its age-sex groups in each county (5-year age group). The expected numbers of deaths is computed using indirect standardization. The standardized mortality ratio ($SMR=\frac{Y}{E}$) is calculated and summarized in Table1. %ADD MORE HERE

Health factors are collected from the publicly available County Health Rankings (CHRs) data released in 2020. %The CHRs highlights health outcomes and health determinants as two crucial parts of the modeling framework. 
The four latent health factors that we aim to construct include: (1) health behaviors, (2) clinical care, (3) social and economic factors, and (4) the physical environment. %Health outcomes include length of life (premature mortality) and quality of life (poor or fair health, poor physical health, poor mental health, and low birth weight). Premature mortality was chosen as the health outcome in this study because few studies have looked at its relationship with racialized economic segregation at the county level.  
%Each health factor is computed based on its associated health indicators, which are documented in the supplementary materials.
Each latent health factor is constructed based on its associated indicators, which are documented in the supplementary materials. Health indicators are standardized to create z-scores for consistency by using the formula: $Z=\frac{\textrm{county value-average value of counties in the US}}{\textrm{standard deviation of counties in the US}}$, where positive z-scores indicate values greater than average in the US and vice versa. In the CHRs 2020 data, the most frequent missing data are observed for the mental health provider measure, with 7.4$\%$ of counties with missing information. We impute missing data using state-level values, based on a method that was proposed by the CHRs and similar to previous studies \citep{cchr,tabb2018assessing}.

Data on the Index of concentration at the Extremes (ICE) are obtained from the Public Health Geocoding Project Monograph for every county in the US (Krieger et al, \citeyear{covid19}). In this study, we focus on racialized economic segregation (ICE$_{race+income}$), which is defined as $ICE_{i}=\frac{A_{i}-P_{i}}{T_{i}}$, where $A_{i}$ represents the number of non-Hispanic White residents in the $80th$ income percentile in county $i$, $P_{i}$ is the number of non-Hispanic Black residents in the $20th$ income percentile in county $i$, and $T_{i}$ is the total population in county $i$ with known income. This measure ranges from -1 to 1, where negative one indicates that all the population in the given county is concentrated in the most deprived group; positive one indicates that all the population is concentrated in the most privileged group; and zero indicates that the deprived and privileged are balanced. The US Census American Community Survey (ACS) 5-year annual average values are used to compute the ICE measure (2014-2018) – which closely aligns with the premature mortality data in the CDC WONDER, based on 2014-2018 data.

\label{sec:data}
\section{Methods}
\label{sec:meth}
The two-stage Bayesian hierarchical statistical framework that we proposed is based on combing the spatial latent factor and the spatially varying coefficient models together, investigating the association between health outcomes and racialized economic segregation as well as the latent health factors. This hierarchical framework is fitted stage-by-stage, with the posteriors from the first stage serving as the priors for the second. Gelman (2004) discussed the computational advantages of this two-stage setup. For instance, the first stage may be conceptualized as \emph{acquiring new information}, and the current stage of the model may represent the current state of knowledge. Furthermore, the second stage of the model algorithm could have a direct interpretation with respect to parameters in the model. Additionally, the lack of an iteration between stage 1 and stage 2 might be a desire for our empirical study since we do not want the health outcome (premature mortality) to play a role in constructing the latent health factors. 

%In the first stage, we model and estimate four health latent factors: health behaviors, clinical care, social and economic status, and the physical environment. In the second stage, we examine the spatial association between premature mortality and racialized economic segregation locally and globally, while controlling for the latent health factors. 

\subsection{Stage 1: Spatial latent factor model for CHR health factors}
We first introduce the spatial latent factor model \citep{wang2003generalized}, which is an extension of the traditional factor analysis by incorporating a spatial structure in the latent factor model. In our study, since we model the four latent health factors separately, we use the health behavior factor as an example to illustrate the model formulation.

Let $z_{ip}$ be the standardized score of indicators for health behavior $p=1,...,P$ at county level $i=1,...,N$. The model can be written as $z_{ip}=\alpha_p+\lambda_p\eta_i+\xi_{ip}$, where $\alpha_p$ is the intercept for the $pth$ indicator, $\eta_i$ is the underlying latent health factor (e.g., health behavior) at county $i$, $\lambda_p$ is the factor loading for the $pth$ indicator, and $\xi_{ip}$ is the error, which is assumed to follow a Normal distribution with mean 0 and variance $\sigma^2_p$. To express the model in a matrix form, lets define $\boldsymbol{z_i}=(z_{1i},...,z_{Pi})^{'}$ to be a vector of $P$ observed indicators and let $\boldsymbol{z}$ be the $NP\times1$ stacked vector of indicators. Then define $\boldsymbol\alpha=(\alpha_{1},...,\alpha_{P})^{'}$ and  $\boldsymbol{\lambda}=(\lambda_{1},...,\lambda_{P})^{'}$ to be the $P\times1$ vector of intercept and factor loadings, respectively. Next, let $\boldsymbol{\eta}=(\eta_{1},...,\eta_{N})^{'}$ to be the $N\times1$ vector of latent factor scores for each county since in our application to the CHRs data, we focus on the case of a univariate latent health variable. Continuing on, let $\boldsymbol{\xi_{p}}=(\xi_{1p},...,\xi_{Np})^{'}$ be the vector of error terms for each indicator, where $\boldsymbol{\xi_p}$ follows a Normal distribution such that $N\sim(\boldsymbol{0_N},\sigma^2_{p}\boldsymbol{I_N})$, where $\boldsymbol{I_N}$ is the identity matrix with dimension $N$ and $\sigma^2_{p}$ measures the residual variation in $\boldsymbol{z_{i}}$. Lastly, let $\sum$ be a diagonal matrix with $(\sigma^2_1,...,\sigma^2_P)^{'}$ along the diagonal and $\sum^*$ be an $NP\times NP$ diagonal matrix such that $\sum^*=\sum\otimes\boldsymbol I_N$, where $\otimes$ is the Kronecker product. To express the model in a more convenient way, define $\Lambda=I_n\otimes \boldsymbol\lambda$ to be a $NP\times N$ matrix of factor loadings. Then under the matrix form, the equation can be written as $\boldsymbol z=\boldsymbol\alpha+\Lambda \boldsymbol \eta+\boldsymbol \xi$, where $\boldsymbol \xi \sim N(0_{NP}, \boldsymbol {\sum^*})$. 

\subsubsection{Likelihood and Prior Specifications}
The spatial factor model can be defined in a hierarchical framework with three levels \citep{hogan2004bayesian}. The first level is the data likelihood. In our study, for the four health latent factors, a normal distribution is assigned since all indicators are continuous and mean scaled variables. Level 1 is specified as: $(\boldsymbol z |\boldsymbol{\alpha,\eta,\Lambda,\sum^*})\sim N(\boldsymbol{\alpha+\Lambda\eta,\sum^*})$. In level 2, we account for the spatial correlation in the latent factor, which specifies the distribution of $\eta_{i}$ conditional on the set
$\eta_{-i}=\eta_{j \ne i} = (\eta_1,...,\eta_{i-1},\eta_{i+1},...,\eta_n)$. In our study, we use the simplest conditional autoregressive (CAR) prior called the intrinsic autoregressive model \citep{besag1991bayesian}, which the conditional distribution of $\eta_i$ is given by $\eta_i|\eta_{-i=j} \sim N(\frac{\sum_jw_{ij}\eta_j}{w_{i+}},\frac{\sigma^2_{\eta}}{w_{i+}})$, where $w_{i+}=\sum_{-i=j}w_{ij}$, and $w_{ij}$ is an $N$x$N$ weight matrix which captures the spatial proximity structure. The most common approach is to assume areas $(i\sim j)$ are neighbors, $i.e.$ $w_{ij}=1$ if and only if $i$ and $j$ share a common border. The prior joint distribution based on the above specification can be written as $p(\eta_i|\sigma^2_{\eta})\propto \sigma^{-N}exp[-\frac{1}{2}\sigma^{-2}_{\eta}\sum_{i=1}\sum_{j<i}w_{ij}(\eta_{i}-\eta_{j})^2]$. It is worth to mention that this prior is improper, since the joint distribution is unchanged even if adding any constant to $\eta$. This further indicates that it cannot model the data directly \citep{abellan2007bayesian}, but it is often used as a prior distribution for random effects. To assure the propriety of the posterior distribution, a constraint such as, $\sum_{j}\eta_{j}=0$ needs to be defined in the model \citep{banerjee2003hierarchical}. Lastly, in level 3, we assign prior distributions for the remaining parameters: the intercept for each indicator $\alpha_p$, the factor loadings for each indicator $\lambda_p$, variance for $\sigma^2_{p}$, and variance for $\sigma^2_{\eta}$. The priors for these parameters are as following: $\alpha_p \sim N(0,a_{\alpha})$, $\lambda_{p=1}=1$, $\lambda_{2-p}\sim N(0,b_{\lambda_{2-p}})$, $\sigma^2_{p}\sim IG(\alpha,\beta)$, $\sigma^2_{\eta}=1$ (to avoid identifiability problem, followed by Abellan et al., 2007). The fixed values of hyperparameters are chosen as follows: $a_{\alpha}=1000$, $b_{\lambda_{2-p}}=1000$, $\alpha=0.5$, and $\beta=0.0005$, respectively. These values are selected to provide non-informative yet proper prior distributions to make the statistical inference more data-driven. Posterior parameter estimates are obtained from samples that are drawn from a Bayesian hierarchical model via Markov Chain Monte Carlo (MCMC).  We run two chains of length 100,000 each and discarded the first 40,000 as burn-in, keeping every 50th to remove the autocorrelation in the simulations. Posterior means and 95\% credible intervals for the factor loadings and the quintile maps for latent health factors are obtained and created. All models are fit in R, via R-NIMBLE \citep{de2017programming} packages. To assess convergence of our MCMC algorithm, we examine trace plots as well as the Gelman-Rubin convergence diagnostics. 

\subsection{Stage 2: Spatially varying coefficient model}
In the second stage model, we aim to assess the association between premature mortality and racialized economic segregation locally and globally, as well as incorporating the estimated latent factors from Stage 1. 

Let $Y_i$ represent the observed premature mortality cases for each county $i=1,...,N$. Then we assume the following spatially varying coefficient model for $Y_i$ follows a Poisson distribution
\begin{center} 
$Y_{i}\given\mu_i\sim Poisson(\mu_iE_i)$
\end{center}

The relative risk is decomposed into the following components
\begin{center} 
$log(\mu_i)=\beta_0+\beta_1X_i+(X_i\delta_i)+\sum_{m=1}^M{\beta_m\eta_{im}}+v_i+\varphi_i$  
%v_i+\varphi_i
\end{center}

where $\beta_0$ is the overall fixed intercept, $\beta_1$ is the global fixed effect of racialized economic segregation, and $\delta_i$ is the spatially varying term which captures the varying effect of racialized economic segregation. $\beta_m$ represents the coefficients for the estimated latent factor $\eta_{im}$, $m=1,...,4$. This model also includes the often-called convolution priors, which has two components: $v_i$ and $\varphi_i$, representing spatially structured and unstructured random effects, respectively. The spatially structured random effects $v_i$ are assigned the intrinsic conditional autoregressive (ICAR) priors, and the unstructured random effects are assumed to follow a Normal distribution $\varphi_i\sim N(0,\sigma^2_{\varphi})$. 

%where $\beta_0$ is the overall fixed intercept, which can be considered as the overall average value of the outcome measure. $\beta_1$ is the global fixed effect of racialized economic segregation, and $\delta_i$ is the spatially varying effect which captures the varying effect of racialized economic segregation. $\beta_m$ represents the coefficients for the estimated latent factor $\eta_{im}, m=1,...4$. This model also includes the often-called convolution priors, which has two components: $v_i$ and $\varphi_i$, representing spatially structured and unstructured random effects, respectively. The spatially structured random effects $v_i$ are assigned the intrinsic conditional autoregressive (ICAR) priors, and the unstructured random effects are assumed to follow a Normal distribution $\varphi_i\sim N(0,\sigma_{\varphi})$. 

We adopt the Integrated Nested Laplace Approximation (INLA) method to estimate the parameters for the above model. INLA is an alternative approach for fitting Bayesian models, based on Laplace approximation, which has been demonstrated as a computationally efficient method compared to traditional Markov Chain Monte Carlo (MCMC) \citep{rue2009approximate, carroll2015comparing}. Under the INLA framework, models are classified as latent Gaussian models, which are applicable in various settings, particularly in spatially varying coefficient model settings \citep{bakka2018spatial}. Let
%\begin{center}
$\boldsymbol{R_1}=\{\boldsymbol{\mu,\beta,\delta,v,\varphi}\}$ be the vector of a Gaussian latent random field and let $\boldsymbol{R_2}=\{\sigma^2_{\delta},\sigma^2_{v},\sigma^2_{\varphi}\}$
%\end{center}
be the vector of hyper-parameters, which are not necessarily Gaussian. Here we assume $\boldsymbol{R_1}$ follows a multivariate Normal distribution $R_1\sim MVN(0,\boldsymbol{Q}^{-1})$ with density $\pi(R_1|R_2)=(2\pi)^{-n/2}|\boldsymbol{Q}|^{1/2}exp\{-\frac{1}{2}R_1^T\boldsymbol{Q}R_1\}$ where $Q$ is the sparse precision matrix (Havard et al.,2009). Then the joint posterior distribution of $\boldsymbol R_1$ and $\boldsymbol R_2$ given the data is: 
\begin{center}
$\pi(\boldsymbol{R_1,R_2|y})=\frac{\pi(\boldsymbol{R_1,R_2,y})}{\pi(\boldsymbol y)}=\frac{\pi(\boldsymbol{y|R_1,R_2})\pi(\boldsymbol{R_1|R_2})\pi(\boldsymbol{R_2})}{\int_{\boldsymbol R_1}^{}\int_{\boldsymbol{R_2}}\pi(\boldsymbol{y|R_1,R_2})\pi(\boldsymbol{R_1|R_2})\pi(\boldsymbol{R_2})d\boldsymbol R_1 d\boldsymbol R_2}$
\end{center}
We could also rewrite the above equation as $\pi(\boldsymbol {R_1,R_2|y}) \propto\pi(\boldsymbol{y|R_1,R_2})\times \pi(\boldsymbol{R_1|R_2})\times\pi(\boldsymbol{R_2})$. Using Laplace approximation, the approximation to $\pi(\boldsymbol{R_2|y})$ using Gaussian distributions can be constructed:
\begin{center}
 ${\tilde\pi}(\boldsymbol{R_2|y})\propto\left. \frac{\pi(\boldsymbol{y|R_1,R_2})\pi(\boldsymbol{R_1|R_2})\pi(\boldsymbol R_2)}{\tilde\pi(\boldsymbol{R_1|R_2,y})}\right\vert_{\boldsymbol{R_1=R_1^*(R_2)}}$
\end{center}
where $\tilde\pi(\boldsymbol{R_1|R_2,y})$ is the Gaussian approximation to the full conditional of $\boldsymbol{R_1}$, and $\boldsymbol{R_1^*(R_2)}$ is the mode of $\boldsymbol R_1$ for a given  configuration of $\boldsymbol R_2$. Then a Laplace approximation approach is used to get 
 $\pi(\boldsymbol{R_1}|\boldsymbol{y})$.
\begin{center}
$\tilde\pi(\boldsymbol{R_{1}|y})\propto\left. \frac{\pi(\boldsymbol{R_1,R_2|y})}{\tilde\pi(\boldsymbol{ R_{-i1}|R_{i1},R_2,y)}}\right\vert_{\boldsymbol{R_{-i1}=R_{-i1}^*(R_1,R_2)}}$
\end{center}
where $\tilde\pi(\boldsymbol{R_{-i1}|R_{i1},R_2,y)}$ is the Gaussian approximation to $\boldsymbol{R_{-i1}|R_{i1}}$, $\boldsymbol{R_2}$, $\boldsymbol{y}$ and\\ $\boldsymbol{R_{-i1}=R_{-i1}^*(R_1,R_2)}$ is the mode. 

\subsubsection{Stage 2 model implementation}
To assess the spatial association between premature mortality and the racialized economic segregation, we fit a series of sequential models. We specify $\varphi_i\sim N(0,\sigma_{\varphi}^2)$ for the unstructured effects, $v_i\sim N(\frac{\sum_jw_{ij}v_j}{w_{i+}},\frac{\sigma^2_{v}}{w_{i+}})$
for the spatial effects and $
\delta_i\sim N(\frac{\sum_jw_{ij}\delta_j}{w_{i+}},\frac{\sigma^2_{\delta}}{w_{i+}})$ for the spatially varying coefficient term for ICE$_{race+income}$ (hereafter ICE),  where $w_{ij}$ is a weight matrix which captures the spatial proximity structure. The most common approach is to assume areas $(i\sim j)$ are neighbors, $i.e.$ $w_{ij}=1$ if and only if $i$ and $j$ share a common border.
We first fit a naïve simple linear regression model to study the overall association between premature mortality and racialized economic segregation (M1). Then, we include the latent health factors based on Stage 1 to assess how these constructed latent factors impact premature mortality (M2). Since premature mortality has been shown to vary geographically, we include a spatial random effect and an unstructured/non-spatial random effect to capture the spatial correlation and residual remaining in the data (M3).  Lastly, we include a spatially varying coefficient term of racialized economic segregation based on our assumption that the measure, operationalized as ICE, can vary based on county (M4).

\begin{flushleft}
Model 1: Naïve model (with main effect of ICE).
\end{flushleft}
\begin{center}
    $log(\mu_i)=\beta_0+\beta_1ICE_i$
\end{center}
Model 2: Model 1 plus CHR conceptual framework for latent health factors.
\begin{center}
$log(\mu_i)=\beta_0+\beta_1ICE_i+\sum_{m=2}^{M}\beta_m\eta_{im}$
\end{center}
Model 3: Model 2 plus spatial effects.
\begin{center}
$log(\mu_i)=\beta_0+\beta_1ICE_i+\sum_{m=2}^{M}\beta_m\eta_{im}+v_i+\varphi_i$
\end{center}
%v_i+\varphi_i
Model 4: Model 3 plus spatially varying effect of ICE.
\begin{center}
$log(\mu_i)=\beta_0+\beta_1ICE_i+\sum_{m=2}^{M}\beta_m\eta_{im}+v_i+\varphi_i+(\delta_iICE_i)$
\end{center}

We assigned a non-informative Gaussian prior distribution for the fixed effects $\boldsymbol \beta\sim N(0,10^3)$. For the hyperparameter precision terms, $\tau=1/\sigma^2$, we assign a weakly-informative Gamma(1,0.5) instead of the commonly used Gamma(1,0.0005) \citep{carroll2015comparing}. Posterior summary statistics, including mean and 95\% credible intervals (CIs) will be presented. To evaluate model performance, we use the Deviance Information Criterion (DIC) and the Wantanabe-Akaike Information Criterion (WAIC) \citep{spiegelhalter2002bayesian,watanabe2010asymptotic}. Both methods have been demonstrated to be effective methods for comparing different models. Smaller DIC and WAIC values indicate a good model performance and efficiency. We fit the above models in R, via R-INLA packages \citep{martins2013bayesian,rue2009approximate}.

%which is a generalization of the Akaike’s information criterion (AIC). The DIC is the sum of model fit $\overline{D}$ and model complexity $p_D$, $DIC=\overline{D}+p_D$ \citep{spiegelhalter2002bayesian}. Smaller DIC indicates a good model performance and efficiency. We fit the above models in R \citep{team2013r}, via R-INLA packages \citep{martins2013bayesian,rue2009approximate}.

%Diagnostic plots for fixed effects as well as random effects will be presented. 

\section{Results}
\label{sec:res}
\subsection{Exploratory analysis}
\begin{figure}[ht]
   \centering 
   \includegraphics[height=6.6cm]{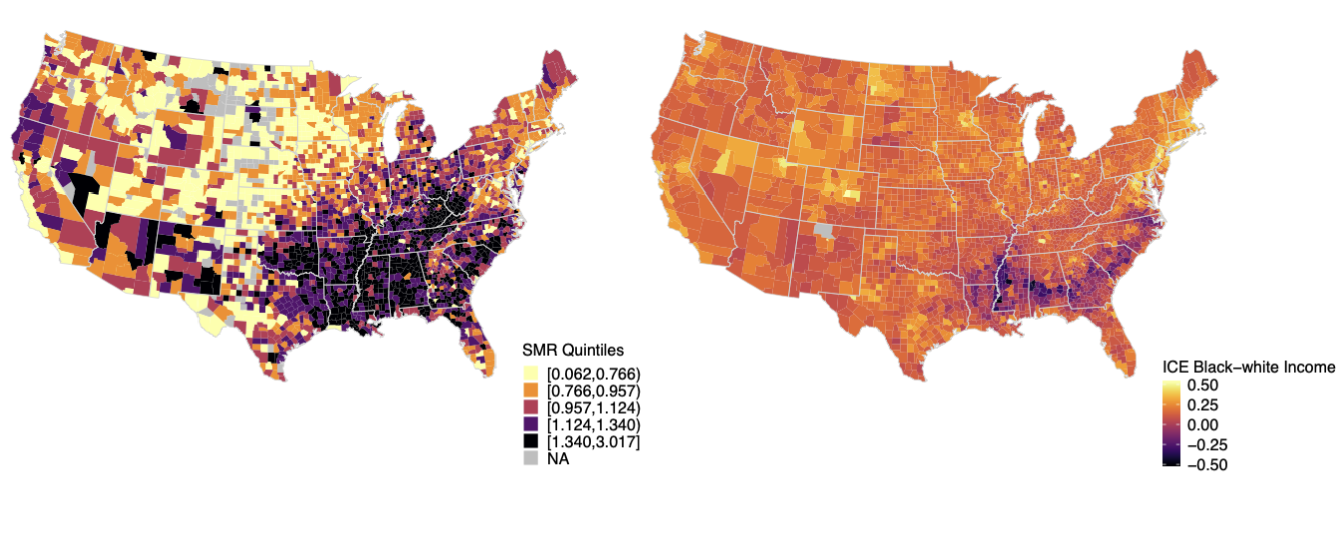}
   \caption{Display of the data. 
   \textbf{(Left)} The standardized mortality ratio (SMR) for premature mortality and 
   \textbf{(Right)} Map of racialized economic segregation at US county level (2014-2018). Negative values indicate higher concentration of low income Black residents and positive values indicate higher concentration of high income White residents. \emph{Note that NA indicates suppressed data. Rio Arriba county in New Mexico has missing data.}}
   \label{fig:ice}
\end{figure}

Results of numeric summary statistics and Moran's I analysis for ICE and health indicators are presented in Table \ref{tab:load1}. The average ICE measure for the 3107 (missing data in Rio Arriba county, New Mexico) US counties is 0.12, the maximum ICE measure is 0.54 and the minimum is -0.52. The ICE measure also yields a highly significant Moran's I statistics equal to $0.65$ $(P<0.001)$, demonstrating the existence of spatial correlation. The Moran's I test is consistent with the US ICE map shown in Figure \ref{fig:ice}, where a clear geographic patterning of the ICE measure is observed. More specifically, we observe a high concentration of low-income Black residents in the southeastern portion of the country, often including many of the \emph{Stroke Belt} states. Evidence of significant spatial correlation $(P<0.001)$ is observed for all health indicators, also indicating the need to incorporate spatial effect to construct the latent health factors. Bivariate correlation test is conducted for all health indicators, where majority of indicators which belong to the same latent health factor are significantly correlated within each other, with few exceptions. We present correlation maps and tables for the heath indicators in the supplementary materials section.

%Results of numeric summary statistics and Moran’s I analysis for all indicators are presented in Table 1. Evidence of significant spatial correlations $(P<0.001)$ was found for all indicators, also indicating the need to use spatial modeling to construct the latent health factors. The ICE measure was also yielded a highly significant I statistics equal to 0.65 $(P<0.001)$, demonstrating the existence of spatial correlations in the racialized economic segregation measure. Figure \ref{fig:ice} presents the raw values of ICE in US counties, where we observe that high concentrations of low-income Black residents are in the southeastern portion of the country, often including many of the \emph{Stroke Belt} states. The geographic patterning is consistent with the Moran’s I test, where spatial correlations are presented in the ICE measure. Bivariate correlation test was conducted for all indicators. Majority of indicators which belong to the same latent health factor were significantly correlated within each other, with few exceptions. We present a correlation maps and tables for the health indicators in the supplementary materials section. 

\subsection{Results for Spatial latent factor model}
%Factor loadings are an important part in factor and spatial factor models. In a spatial factor model, factor loadings can be interpreted as weights or the association between latent health factors and indicators. 

Table 1 presents the posterior means as well as the 95\% credible interval (CrI) of factor loadings for all indicators. In our study, we interpret factor loadings as weights for each indicators. The degree of health behavior, social and economic status, and physical environment are mainly driven by the following indicators: teen birth rate, children in poverty, and long commute-driving alone, respectively. Most clinical care indicators have a similar contribution to the constructed latent factor, except for the preventable hospital stay rate, which show the lowest impact. 
%By applying the spatial factor model to construct these four latent health factors, 
Additionally, the contribution for the indicators is different than the CHR deterministic weights – which are initially proposed in the original formulation of the CHR conceptual framework \citep{remington2015county}. For example, the CHRs assigns more weights on adult smoking (health behaviors), uninsured and preventable hospital stays (clinical care), unemployment (social and economic factors), and air pollution (physical environment) for the four health factors, whereas our model shows that the weights of the highest magnitudes for these health factors are: teen birth rate, children in poverty, primary care physicians, and long commute-driving alone.

\begin{figure}[ht]
   \centering
   \includegraphics[width=1\textwidth,height=12cm]{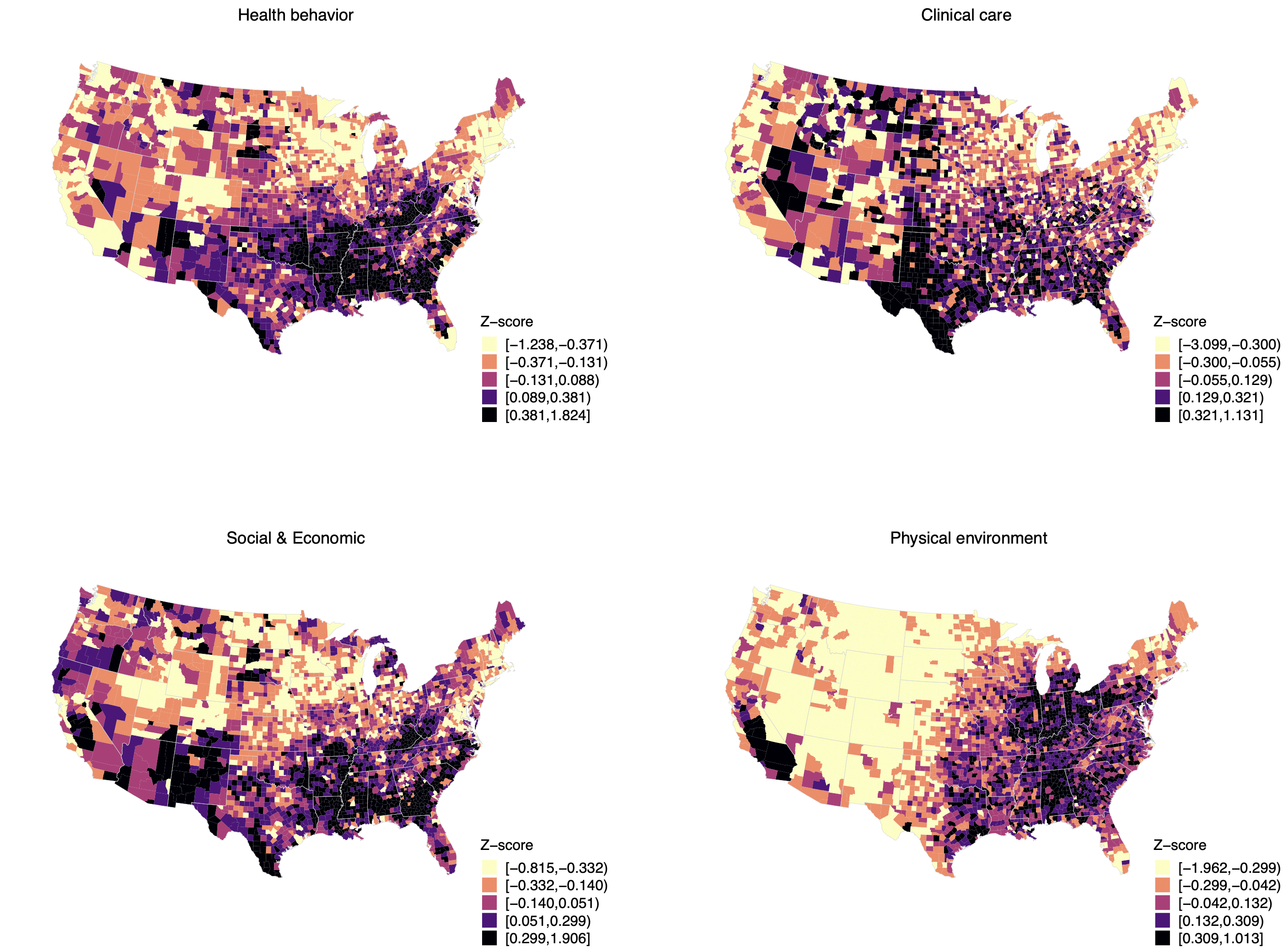}
   \caption{Quintile maps of latent health factors (positive scores indicate poor health factor and negative scores indicate ideal health factor)}
   \label{fig:latent_factor}
\end{figure}

To further demonstrate the spatial and geographic patterning of the health factors, we provide maps of the four latent health factors (Figure \ref{fig:latent_factor}). These maps clearly reflect the spatial patterning for all latent health factors in the United States, where dark purple represents areas with poor health behavior, clinical care, social-economic and physical environment status, and light colors indicate areas with ideal health factors. More specifically, states in the southeastern portion of the US, many that fall in the \emph{Stroke Belt} have larger numbers of counties with poor health behavior status compared to states in the northern, or even western region of the US. Areas with poor clinical care status are concentrated in Texas, Nevada and the majority of \emph{Stroke Belt}. States in the south – both the southeast and the southwest – have more counties with poor social and economic status compared to states in the north. Additionally, California and the eastern part of the US have more counties with a poor physical environment quality. In contrast, most states in the western part of the US have satisfactory physical environmental status.
%\begin{figure}[h]
%    \centering
 %   \includegraphics[width=0.25\textwidth]{mesh}
%    \caption{a nice plot}
%    \label{fig:mesh1}
%\end{figure}

%We present the exceedance probabilities of counties falling into the least health factor quintiles for each health factor; more specifically, we examined the posterior distribution of latent health factors for counties exceeding the $80th$ percentile. 

\begin{figure}[ht]
\centering
   \includegraphics[width=1\textwidth,height=10cm]{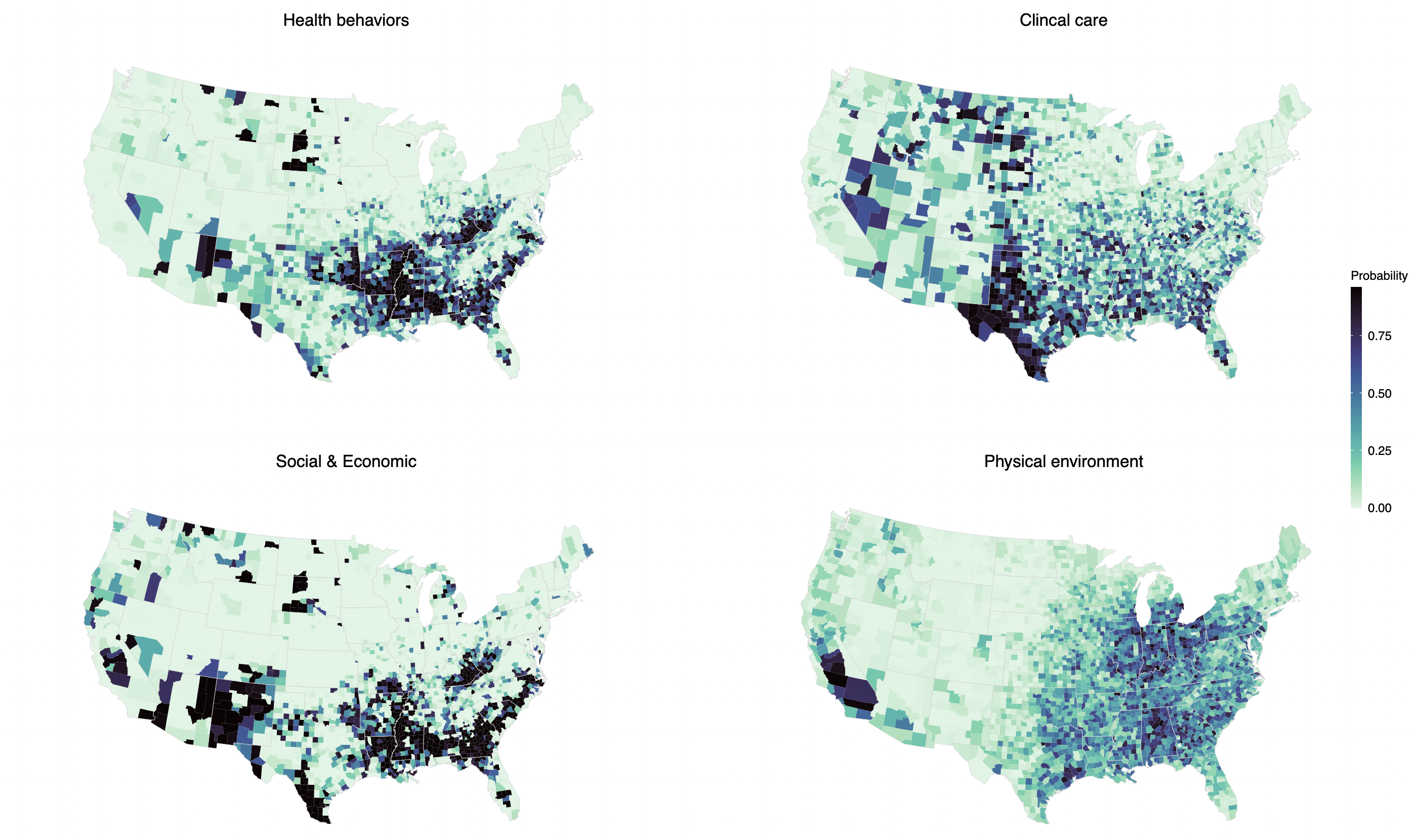}
   \caption{Probability maps for latent health factors}
   \label{fig:prob}
\end{figure}

Figure \ref{fig:prob} shows the exceedance probabilities of counties falling into the least health factor quintiles (exceeding the $80th$ percentile) for health factors. Many counties in the \emph{Stoke Belt} states are most likely to have unhealthy behaviors. Counties in Texas, particularly along the Mexico border, have an 80\% higher chance of being in the worst clinical care conditions. Additionally, southern counties are more likely to be socially and economically impoverished. We also find that many eastern US counties are more likely to have disadvantaged physical environment than western US counties, with the exception of California, in which several counties, including Madera, Fresno, Kings, Tulare, Kern, Los Angeles, San Bernardino, and Riverside, have noticeable poor physical environment.

\subsection{Results for spatially varying coefficient models}
Table 2 presents the results of the association between premature mortality, racialized economic segregation (as measured by ICE), and the latent health factors. Due to the multicollinearity issue between ICE and social $\&$ economic factors, we exclude it from the model. Posterior means of the regression effects for the fixed effects as well as 95\% Bayesian credible interval are provided for each parameter. DIC and WAIC values of each model and summaries of the hyper-parameters are also provided. Across the four models, there is a negative significant association between premature mortality and ICE in Models 1 through 4. This suggests that in areas of high concentration of low income Black residents, the average premature mortality risk is higher. The parameter estimates of ICE are attenuated by 82.7\% after increasing the model complexity (comparing Model 1 to Model 4). The parameter estimates for latent health factors are consistent across Models 2, 3, and 4, after controlling for other variables. Specifically, there is a significant positive association between premature mortality and health behavior, indicating that the risk of premature mortality increased as unhealthy behavior scores increased. Physical environment is also significantly associated with premature mortality, implying that a poor physical environment may contribute to an increased risk of premature death. Clinical care, on the other hand, is negatively associated with premature mortality. 

From Model 1 to Model 4, a clear improvement in both DIC and WAIC values are observed, with Model 4 showing the best model fit, with the lowest DIC and WAIC values. There is a strong evidence that the null model (Model 1) cannot adequately describe the relationship between premature mortality and racialized economic segregation at US county level. When comparing Models 2 and 3, we observe a significant drop in the DIC and WAIC values by adding on structured and unstructured random effects. In addition, a decrease in the DIC and WAIC values have also been observed when adding the spatially varying ICE random effect in model 4. The precision terms for the structured and unstructured county level random effects are similar between Models 3 and 4, with model 4 having slightly larger precision terms. Given the importance of incorporating neighborhood level latent health factors, the spatially varying nature of the racialized economic segregation, and the objective evidence based on DIC and WAIC values, we will focus on interpreting Model 4 for the remainder of the paper. Model 4 shows that there is a significant negative  association between ICE and premature mortality after controlling for other latent health factors, more specifically, an increase of 1 unit in the ICE measure is associated with a decrease of 24\% in the risk of premature mortality $(e^{\beta}=0.757,$ $95\%$ credible interval: $0.673, 0.852)$. This further indicates that people living in areas with higher concentrations of low-income Black residents have, on average, a higher risk of premature mortality. Neighborhood level latent health factors are all significantly associated with premature mortality, after controlling for the fixed ICE effect. Health behaviors and physical environment are positive associated with premature mortality $(e^{\beta}=1.633,$ $95\%$ credible interval: $1.576, 1.695;$ $e^{\beta}=1.406,$ $95\%$ credible interval: $1.368, 1.444)$. These results suggest that the average risk of premature mortality is higher in areas with unhealthy behaviors and a poor physical environment. However, we observe a significant negative association between clinical care and premature mortality $(e^{\beta}=0.794,$ $95\%$ credible interval: $0.771, 0.817)$, indicating that areas with poor clinical care conditions may not have higher risk of premature mortality. For instance, based on Figures \ref{fig:latent_factor} and \ref{fig:prob}, we observe that Texas has the worst clinical care system when compared to other states; however, Figure \ref{fig:ice} shows that many counties in Texas do not have higher risk of premature mortality. 

%Given that there are 121 counties with suppressed data in our study, more research should be done to assess the relationship between premature mortality risk and clinical care in US counties.

\begin{figure}[ht]
   \centering   \includegraphics[width=1\textwidth,height=12cm]{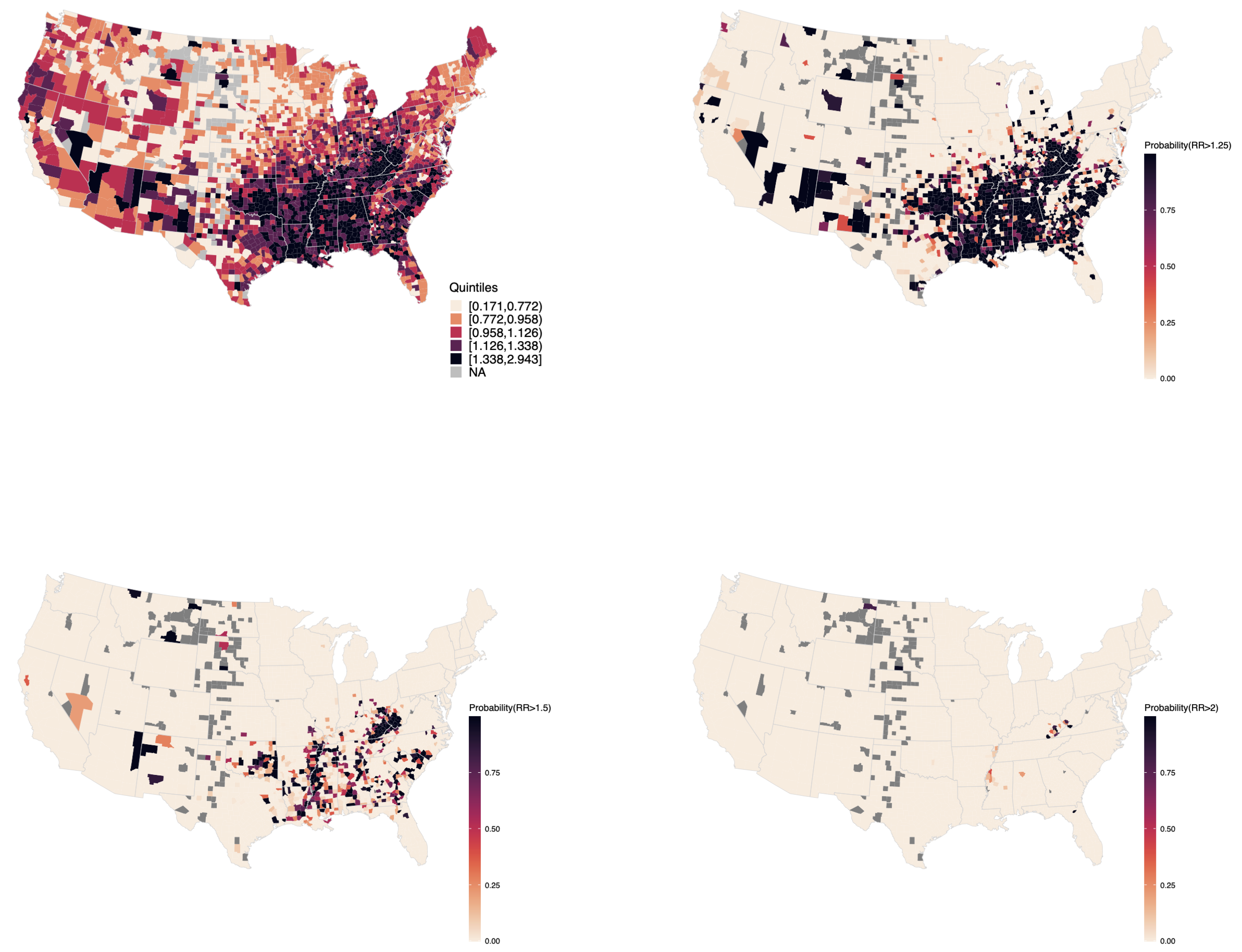}
   \caption{Maps of relative risk and exceedance probabilities with different thresholds (1.25, 1.5, and 2) for premature mortality at US county level (2014-2018). \emph{Note that counties in grey are due to suppressed data.}}
   \label{fig:smrm}
\end{figure}

Figure \ref{fig:smrm} depicts the estimated relative risk $(e^\mu)$ of premature mortality for US counties based on model 4. This map not only depicts the spatial distribution of premature mortality risk, but it also captures a concentration of elevated premature mortality risk in the Southeast portion of the United States, as indicated by relative risks greater than one (darker purple in color). Furthermore, many counties in the Northeast states, such as West Virginia and Ohio, have a higher risk of premature death.

To detect unusual elevations in disease risk based on the different thresholds, we compute the exceedance probabilities of relative risk estimates of premature mortality greater than 1.25, 1.5, and 2. Counties with probabilities close to 1 are very likely to have relative risks exceed thresholds, while counties with probabilities close to 0 are unlikely to have relative risks exceed thresholds. From Figure \ref{fig:smrm}, we observe that the relative risk of premature mortality in most of the \emph{Stroke Belt} counties has higher probabilities greater than 1.25 and 1.5. Whereas few counties in the US have higher probabilities that the premature mortality risk is greater than 2. 

\begin{figure}[ht]
   \centering
   \includegraphics[width=0.8\textwidth,height=9cm]{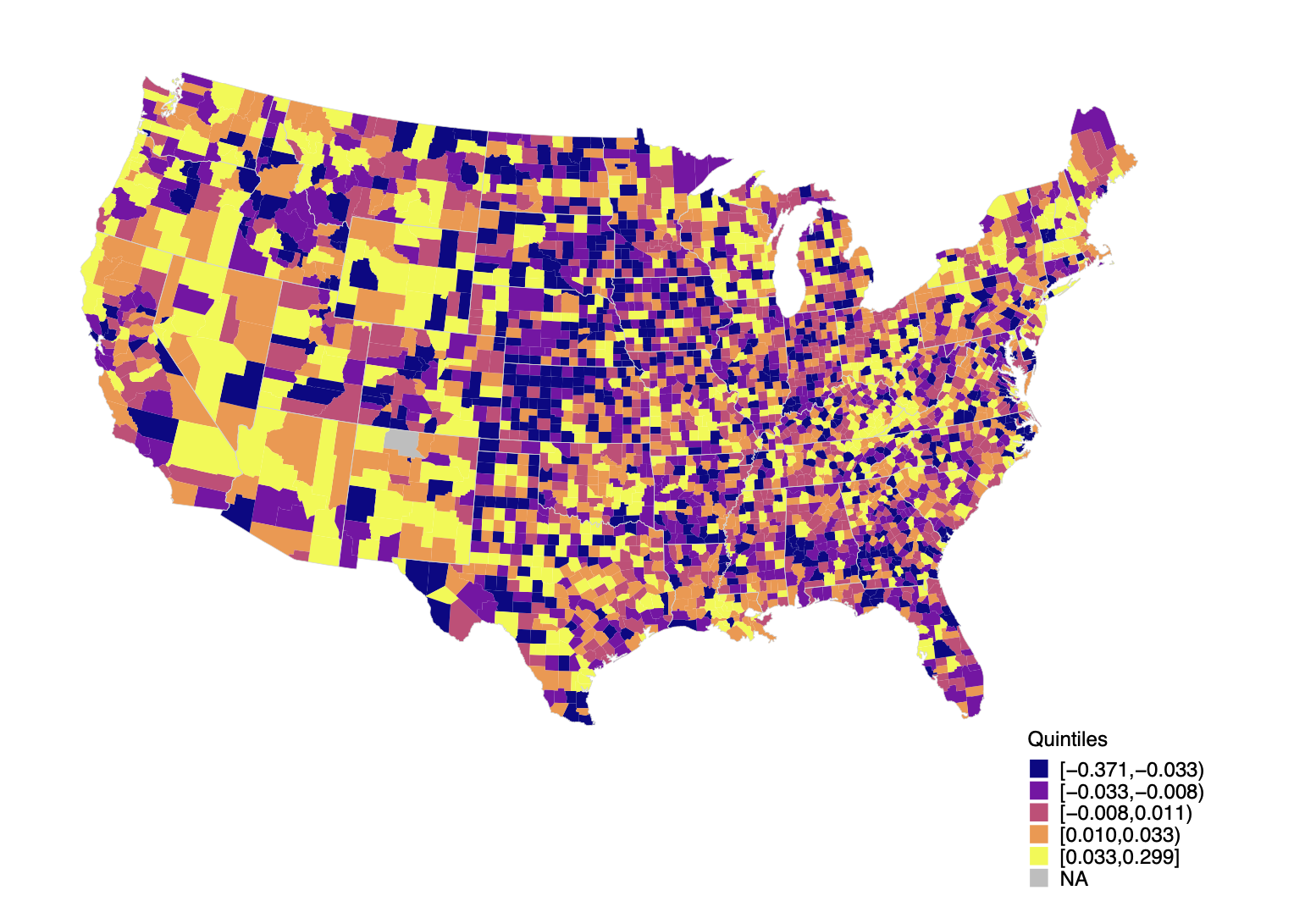}
   \caption{Map for varying ICE$_{race+income} (\delta_{i}$) (2014-2018). \emph{Note that Rio Arriba county in New Mexico has missing data.}}
   \label{fig:vary}
\end{figure}

Lastly, in Figure \ref{fig:vary}, we present the quintile map of ICE random effect based on Model 4. This map suggests that ICE in premature mortality vary across the US, at the county level - where darker shaded counties indicate more negative random effects and lighter shaded counties indicate positive random effects. Moreover, we also observe that the ICE random effects are locally clustered, with neighboring counties having similar random effects. Additionally, as evidenced by the relatively small precision term $(\tau_{\delta}=7.25)$ in Model 4, there may be other unknown information that has not been fully captured by the varying effects.

%We also compute the exceedance probabilities of relative risk estimates of premature mortality greater than 1.5 in order to detect unusual elevations in disease risk. Figure \ref{fig:smrm} shows the map of the exceedance probabilities for premature mortality in US counties. Specifically, counties with probabilities close to 1 are very likely to have relative risks that exceed 1.5, while counties with probabilities close to 0 are unlikely to have relative risks that exceed 1.5. The map shows that many counties with a higher probability of elevated premature mortality risk fall in the \emph{Stoke Belt}. In addition, some counties in Oklahoma and along the Texas border have an elevated risk of premature death. %NEED TO JUSTIFY THE LANGUAGE%
%\begin{figure}[h]
%   \centering
   %\includegraphics[width=0.8\textwidth,height=9cm]{probpm.png}
 %  \caption{Map of exceedance probabilities for premature mortality at US county level (2014-2018). \emph{Note that counties in grey are due to suppressed data.}}
 %  \label{fig:probpm}
%\end{figure}

\section{Discussion}
%This paper proposed a two-stage Bayesian modeling framework, where combined and illustrated the application of the spatial latent factor model and spatially varying coefficient model to assess the association between premature mortality and racialized economic segregation while adjusting the neighborhood-level latent health factors. Assessing the relationship between racialized economic segregation and health outcomes is essential in both policy and research. It has been extensively studied in social epidemiology; however, taking a step further and considering spatially varying racialized economic segregation effect in the model is critical for revealing the underlying relationship between them. Although any measure of a latent health construct is bound to be imperfect, our model-based approach provides a method for gaining available information from spatially correlated indicators, adequately reducing the dimension of the data without losing information. 
In this paper, we have proposed a two-stage Bayesian modeling framework, where combine and illustrate the application of the spatial latent factor model and spatially varying coefficient model to assess the association between premature mortality and racialized economic segregation while adjusting the neighborhood-level latent health factors. The model framework we propose bridges the gap between spatial latent factor models for spatially correlated health indicators and spatially varying coefficient models, which are difficult to fit when covariates are large. Specifically, this framework can accommodate different types of latent factors at stage 1, such as the deprivation index. At the second stage model, other segregation indices which are considered spatial could also be studied at the global and local levels, with various health outcomes. As a result, a wide range of applications from previous literature (at the area level) may use our framework and thus achieve research goals.

%Our proposed modelling framework provides a guideline to study 

Overall, This analysis shows that racialized economic segregation has a significant impact on premature mortality; counties with a high concentration of low-income Black residents have, on average, a higher risk of premature mortality than counties with a high concentration of high-income White residents. Although attenuation is evident, we also find that this impact persists after adjusting for latent health factors. Specifically, health behavior, clinical care, and physical environment are significantly contributed to premature mortality risk. These findings also suggest that the risk of premature death is higher in areas with unhealthy behaviors and a poor physical environment. In addition, our maps reveal the apparent geographical patterning of racialized economic segregation and latent health factors, which is aligned with our initial hypothesis of these variables.

Our findings of the association of racialized economic segregation with premature mortality are consistent with previous works. One study assessed the relationship between premature mortality and racialized economic segregation among Black and white adults in Boston \citep{krieger2017measures}. Another study based on Black and white adults examined the association between premature mortality and racialized economic segregation in Chicago \citep{lange2018association}. Both studies found strong significant relationships between premature mortality and racialized economic segregation. These studies assumed racialized economic segregation as a fixed effect (i.e., global effect) in their models and did not consider the spatial nature of this measure. Additionally, our results are also consistent with a previous study that have examined the spatial association between overall health and health factors, where a strong significant association between overall health and health behavior was observed by using a geographically weighted regression approach, although health factors were constructed based on the CHRs pre-defined weights (Tabb et al., 2018).

One novel aspect of our work is that we consider the spatial nature of racialized economic segregation, as opposed to treating it as a fixed effect in the model. Our study explained the spatial patterns and the varying effects of racialized economic segregation on premature mortality by utilizing spatially varying coefficient models, in contrast to prior studies. Furthermore, we constructed latent health factors based on advanced Bayesian spatial latent factor models instead of only using the CHRs pre-defined weights to calculate these factors. Several studies have been carried out to investigate the relationship between health outcomes and health factors, with health factors calculated using the CHRs approach (Peppard et al., 2008; Anderson et al., 2015; Remington et al., 2016; Greer et al., 2016; Niazi et al., 2021). However, as argued by other studies, the CHRs weights neither consider spatial correlation among indicators nor use an advanced modeling approach, which could result in biased estimates when constructing health factors (Courtemanche et al., 2015; Davis et al., 2021). Although any measure of a latent health factor is bound to be imperfect, our model-based approach highlighted the ability of spatial factor analysis to reduce the dimensionality of spatially correlated datasets without losing information when estimating latent health factors.

Some limitations in this study are worth mentioning. Firstly, 121 counties have suppressed premature mortality data (2014-2018) due to the CDC WONDER policy. Thus, we treat these counties as having missing data. 
Secondly, some measures from the CHRs data are based on previous years. For example, for the health behaviors factor, the adult smoking indicator is based on 2017 data; however, the teen birth indicator is based on 2012-2018 data. Additionally, there are some missing data shown for some CHR measures. For instance, violent crime rates were missing for 229 counties (229/3108*100=7.4\%); thus, we used the state average violent crime rates to impute these missing values. Thirdly, this study relied on county-level aggregated data and did not include individual-level covariates; thus, our results might suffer from ecological bias. Lastly, it is possible that unmeasured confounding variables may play a role in estimating the varying racialized economic segregation in premature mortality across the US counties.

In conclusion, our analysis constructed four latent health factors, and proved the presence of spatial correlation in these health factors. We have mapped and measured the racialized economic segregation effect in premature mortality in the US and showed that both racialized economic segregation and latent health factors play a significant role in estimating premature mortality. The spatially varying coefficient model adds an advantage over the typical fixed effect model in that it highlights the varying effects of the factors like racialized economic segregation. Using the CHRs data, further study could investigate the association between racialized economic segregation and overall health outcomes (including mortality and morbidity). Additionally, examining the spatial and temporal effects of racialized economic segregation will allow for a complete understanding of its impact on not only premature mortality that we examined in this research, but in many other settings that examine health and social inequities.

%\bigskip
%\begin{center}
%{\large\bf SUPPLEMENTAL MATERIALS}
%\end{center}

%\bibliographystyle{jasa} 

\bibliographystyle{jasa}  
\bibliography{references}

%\newpage 
%\begin{figure}[t]
%   \centering
%   \includegraphics[width=1\textwidth,height=12cm]{LP.png}
 %  \caption{Quintile maps of latent health factors (positive score indicates poor health factor and negative indicates ideal health factor)}
 %  \label{fig:latent_factor}
%\end{figure}

%\begin{figure}[t]
%   \centering
%   \includegraphics[width=1\textwidth,height=10cm]{PP.png}
%   \caption{Probability maps for latent health factors}
%   \label{fig:prob}
%\end{figure}

\clearpage
%\newpage
%\begin{figure}[p]
 %  \centering
  %  \includegraphics[width=1\textwidth,height=12cm] %{varying.png}
 %   \caption{Quintile map for varying $ICE_{rece+income} (\delta_{i}$)}
  %  \label{fig:varying}
 %\end{figure}

\begin{table}[t]
%\begin{singlespace}
\begin{center}
\begin{tabular}{l c c c c c c}
\hline							
\multicolumn{1}{l}{} & Mean & SD & Minimum & Median & Maximum & Moran's I \\
\hline	
\vspace{2mm}
\textbf{Health behavior }\\
Adult smoking (\%) & 0.17 & 0.04 & 0.06 & 0.17 & 0.41 & $0.66^{\ast}$ \\
Adult obesity (\%) & 0.33 & 0.05 &	0.12&	0.33&	0.58& 	$0.38^\ast$\\
Food environment index &7.46 &	1.13 &	0.00 &	7.70 &	10.00 &	$0.47^\ast$\\
Physical inactivity (\%) &0.27&	0.06&	0.10&	0.27&	0.50&	$0.48^\ast$\\
Access to exercise opportunities (\%)&	0.62&	0.23&	0.00&	0.66&	1.00&	$0.34^\ast$\\
Excessive drinking (\%)	&0.17	&0.03&	0.08&	0.18&	0.29&	$0.73^\ast$\\
Alcohol impaired driving deaths (\%)&	0.28&	0.15&	0.00&	0.28&	1.00&	$0.15^\ast$\\
Sexually transmitted infections rate &	396.25&	265.90&	35.80&	327.45&	6120.30 &	$0.31^\ast$\\
\vspace{2mm}
Teen births rate  &	29.78&	13.89&	2.11&	28.16&	103.05&	$0.54^\ast$\\
\vspace{2mm}
\textbf{Clinical care }\\
Uninsured (\%) & 	0.11& 	0.05& 	0.02& 	0.11& 	0.34& 	$0.79^\ast$\\
Primary care physicians’ rate &	54.06&	33.84&	0.00&	48.25&	514.45&	$0.11^\ast$\\
Dentist’s rate & 45.44&	30.92&	0.00&	41.06&	711.34&	$0.13^\ast$\\

Mental health providers rate& 	155.12&	159.82&	0.00&	111.50&	2123.03&	$0.29^\ast$\\
Preventable hospital stays rate& 4863.93& 1831.37&	536.00&	4724.00&	16851.00&	$0.52^\ast$\\
Mammography screening (\%)	&0.41&	0.08&	0.13&	0.41&	0.65&	$0.61^\ast$\\
\vspace{2mm}
Flu vaccinations (\%)&	0.42&	0.10&	0.05&	0.43&	0.66&	$0.45^\ast$\\
\vspace{2mm}
\textbf{Social and economic}\\

High school graduation rate& 0.89&	0.07&	0.26&	0.90&	1.00&	$0.34^\ast$\\
Some college (\%)&	0.58&	0.12&	0.15&	0.58&	1.00&	$0.42^\ast$\\
Unemployment (\%)&	0.04&	0.01&	0.01&	0.04&	0.18&	$0.59^\ast$\\
Children in poverty (\%)&	0.21&	0.09&	0.03&	0.20&	0.68&	$0.59^\ast$\\

Income inequality ratio	&4.52&	0.76&	2.54&	4.40&	11.97&	$0.37^\ast$\\
Children in single parent households&	0.32&	0.11&	0.00&	0.32&	0.87&	$0.42^\ast$\\
Social associations	&11.73&	5.89&	0.00&	11.15&	52.31&	$0.41^\ast$\\
Violent crime rate&	249.71&	187.14&	0.00&	205.45&	1819.51& $0.30^\ast$\\
\vspace{2mm}
Injury deaths rate&	86.68&	24.82&	22.30&	83.82&	319.86&	$0.40^\ast$\\
\vspace{2mm}
\textbf{Physical environment}\\

PM2.5 daily average&	9.02&	1.97&	3.00&	9.40&	19.70&	$0.92^\ast$\\
Severe housing problems (\%)&0.14&	0.04&	0.03&	0.13&	0.45&	$0.47^\ast$\\
Driving alone to work (\%)	&0.80&	0.07&	0.06&	0.81&	0.96&	$0.49^\ast$\\
\vspace{2mm}
Long commute driving alone (\%)	&0.32&	0.12&	0.03&	0.31&	0.83&	$0.36^\ast$\\
\vspace{2mm}
\textbf{Racialized economic segregation}\\
\vspace{2mm}
ICE$_{race+income}$&	0.14&	0.12&	-0.52&	0.15&	0.54&	$0.65^\ast$\\
\vspace{2mm}
\textbf{Health outcome}\\
Premature mortality (SMR) & 1.05 & 0.35 &0.06 & 1.04 &3.02\\
\hline
\end{tabular}
\end{center}
  \caption{Summary statistics of all county-level indicators, ICE measure and health outcome.}
  \label{tab:load1}
  %\end{singlespace}
\end{table}

\clearpage
\begin{table}[t]
\captionsetup{width=16cm}
\begin{singlespace}
\begin{center}
\begin{tabular}{cl cl}
\hline							
Latent variable	&	Indicator	&	Mean (95\% CI)	\\
\hline						
\emph{Health behavior loadings}  \\

  & Adult smoking (\%) &  1 \\
  & Adult obesity (\%) & 1.207 (1.134, 1.277)\\
  & Food environment index & -1.219 (-1.292, -1.143)\\
  & Physical inactivity (\%) & 1.501 (1.432, 1.570) \\
  & Access to exercise opportunities (\%) & -1.028 (-1.103, -0.952)\\
  & Excessive drinking (\%)& -1.367 (-1.438, -1.297)\\
  & Alcohol impaired driving deaths (\%) & -0.200 (-0.280, -0.122)\\
  & Sexually transmitted infections rate & 0.664 (0.583, 0.745)\\
  & Teen births rate & 1.524 (1.457, 1.595)\\

\emph{Clinical care loadings} \\

  &	Uninsured (\%) & 1\\
  & Primary care physicians rate  & -1.446 (-1.533, -1.357)\\
  & Dentist rate  & -1.446 (-1.534, -1.356)\\
  & Mental health providers rate  & -1.124 (-1.212, -1.035)\\
  & Preventable hospital stays rate & 0.645 (0.559, 0.731)\\
  & Mammography screening (\%) & -0.970 (-1.060, -0.876)\\
  & Flu vaccinations (\%) & -0.968 (-1.059, -0.876)\\
  
\emph{Social and economic loadings}\\
 
 & 	Income inequality ratio & 1\\
 & High school graduation rate & -0.529 (-0.613, -0.445)\\
 & Some college (\%) & -1.447 (-1.517, -1.375)\\
 & 	Unemployment (\%) & 1.349 (1.277, 1.421)\\
 & 	Children in poverty (\%) & 2.110 (2.042, 2.171)\\
 &  Children in single parent households (\%) & 1.602 (1.531, 1.671)\\
 &  Social associations & -0.385 (-0.469, -0.304)\\
 &  Violent crime rate & 0.793 (0.711, 0.874)\\
 &  Injury deaths rate & 0.892 (0.813, 0.972)\\
 
\emph{Physical environment loadings}\\

 & Air pollution $PM_{2.5}$ daily average & 1\\
 & Severe housing problems (\%) & -0.008 (-0.111, 0.088)\\
 & Driving alone to work (\%) & -0.109 (-0.215, 0.002)\\
 & Long commute driving alone (\%) & 1.019 (0.899, 1.144)\\
 & Drinking water violation & 0.637 (0.529, 0.741)\\
 
 \hline

\end{tabular}
\end{center}
  \caption{Posterior means and 95 \% credible interval of the factor loadings for spatial factor Bayesian model.}
  \label{tab:load2}
  \end{singlespace}
\end{table}

\clearpage
\begin{table}[t]
%\captionsetup{width=18cm}
\begin{singlespace}
\begin{center}
\begin{tabular}{c c c c c}
\hline							
\multicolumn{1}{c}{} & Model 1 & Model 2 & Model 3 & Model 4\\
\hline	
  & Posterior mean & Posterior mean & Posterior mean & Posterior mean\\
  & (95 \% CI) & ( 95\%  CI) & (95 \%  CI) & (95 \%  CI) \\
  
\hline
\vspace{3mm}
\textbf{Fixed effects} \\
$ICE_{race+income}$ & -1.616 & -0.532 & -0.265 & -0.278\\
 & (-1.623, -1.609) & (-0.543, -0.521) & (-0.382, -0.149) & (-0.396, -0.160)\\

\vspace{3mm}
\textbf{Latent health construct}\\

Health behaviors 
                 & & 0.454 & 0.490 & 0.491\\
                 & &  (0.451, 0.458)&(0.454, 0.526) & (0.455, 0.528)\\

Clinical care 
                 & & -0.151 & -0.225 & -0.230\\
                 & & (-0.154, -0.148) & (-0.254, -0.196)& (-0.259, -0.201)\\

Physical environment  
                      & &  0.099 & 0.339 & 0.341\\
                      & & 	(0.097, 0.101) & (0.312, 0.366) & (0.314, 0.368)\\

\vspace{3mm}
\textbf{Random effects}\\

\vspace{2mm}
$\tau_{\varphi}=1/\sigma^2_{\varphi}$ & &  & 20.49 & 21.03\\
\vspace{2mm}
$\tau_{v}=1/\sigma^2_{v}$ & & & 	64.27 & 64.92\\
\vspace{2mm}
$\tau_{\delta}=1/\sigma^2_{\delta}$  & & & & 7.25\\ 
\vspace{2mm}
\textbf{Model fit}\\

 DIC & 129948.20 & 94473.12 & 31313.51 & \textbf{31302.58} \\
 WAIC &1440516.98 & 734204.24&31021.38 & \textbf{31000.72}\\
\hline

\end{tabular}
\end{center}
  \caption{Results for Bayesian models. $\tau_{\varphi}=1/\sigma^2_{\varphi}$ is the precision for the unstructured county level random effects; $\tau_{v}=1/\sigma^2_{v}$ is the precision for the structured county level random effects; $\tau_{\delta}=1/\sigma^2_{\delta}$ is the precision for the structured county level varying ICE coefficient. }
  \label{tab:load3}
  \end{singlespace}
\end{table}

%\begin{figure}[t]
 %  \centering
%   \includegraphics[width=1\textwidth,height=12cm]{ice4.png}
%    \caption{Index of concentration at the extreme Black-white Income (2018)}
 %  \label{fig:prob}
%\end{figure}

%\clearpage
 %\begin{center}
 %  {\bf APPENDIX}
% \end{center}

\end{document}